\documentclass[fleqn,usenatbib]{mnras}

\usepackage{newtxtext,newtxmath}

\usepackage[T1]{fontenc}
\usepackage{ae,aecompl}
\usepackage{textcomp}
\usepackage{gensymb}
\usepackage{threeparttable}
\usepackage{verbatim}
\usepackage{float} 

\usepackage{graphicx}

\usepackage{graphicx}	
\usepackage{amsmath}	
\usepackage{amssymb}	
\usepackage{natbib}
\usepackage{verbatim}

\title[Constraining planet oblateness]{Constraining the oblateness of transiting planets with photometry and spectroscopy}

\author[B. Akinsanmi et al.]{
B. Akinsanmi,$^{1,2,6,}$\thanks{E-mail: tunde.akinsanmi@astro.up.pt}
S. C. C. Barros,$^{1}$
N. C. Santos,$^{1,2}$
M. Oshagh$^{1,3,4}$
and L. M. Serrano$^{5}$
\\
$^{1}$Instituto de Astrof\'isica e Ci\^encias do Espa\c co, Universidade do Porto, CAUP, Rua das Estrelas, 4150-762 Porto, Portugal.\\
$^{2}$Departamento de F\'isica e Astronomia, Faculdade de Ci\^encias, Universidade do Porto, Rua do Campo Alegre, 4169-007 Porto, Portugal.\\
$^{3}$Institut f\"ur Astrophysik, Georg-August-Universit\"at G\"ottingen, Friedrich-Hund-Platz 1, 37077 G\"ottingen, Germany.\\
$^{4}$Instituto de Astrof\'isica de Canarias (IAC), E-38200 La Laguna, Tenerife, Spain.\\
$^{5}$Dipartimento di Fisica, Universita degli Studi di Torino, via Pietro Giuria 1, I-10125, Torino, Italy.\\
$^{6}$National Space Research and Development Agency. Airport Road, Abuja, Nigeria.
}

\date{Accepted 2020 July 16. Received 2020 July 16; in original form 2020 June 12}

\pubyear{2015}

\begin{document}
\label{firstpage}
\pagerange{\pageref{firstpage}--\pageref{lastpage}}
\maketitle

\begin{abstract}
Rapid planetary rotation can cause the equilibrium shape of a planet to be oblate.  While planetary oblateness has mostly been probed by examining the subtle ingress and egress features in photometric transit light curves, we investigate the effect of oblateness on the spectroscopic Rossiter-McLaughlin (RM) signals. We found that a giant planet, with planet-to-star radius ratio of 0.15 and Saturn-like oblateness  of 0.098, can cause  spectroscopic signatures with amplitudes up to 1.1\,m\,s$^{-1}$ which is detectable by high-precision spectrographs such as \texttt{ESPRESSO}. We also found that the spectroscopic oblateness signals are particularly amplified for transits across rapidly rotating stars and for planets with spin-orbit misalignment thereby making them more prominent than the photometric signals at some transit orientations. We compared the detectability of oblateness in photometry and spectroscopy and found that photometric light curves are more sensitive to detecting oblateness than the   spectroscopic RM signals mostly because they can be sampled with higher cadence to better probe the oblateness ingress and egress anomaly. However, joint analyses of the light curve and RM signal of a transiting planet provides more accurate and precise estimate of the planet's oblateness. Therefore, \texttt{ESPRESSO} alongside  ongoing and upcoming photometric instruments such as \texttt{TESS}, \texttt{CHEOPS}, \texttt{PLATO} and \texttt{JWST} will be extremely useful in measuring planet oblateness.

\end{abstract}




\section{Introduction}
Planets have non-spherical equilibrium shapes as a result of different forces acting upon them such as gravitational, pressure, tidal and centrifugal forces. The centrifugal acceleration caused by planetary rotation reduces the effective  gravitational acceleration at the equator compared to the pole thereby leading to an equatorial bulge. The resulting non-spherical planet shape is referred to as oblateness \citep{seagerhui} and is defined by an oblateness parameter, $f$, given as (e.g., \citealt{barnes03,carterwinna}) 
\begin{equation}
       f=\frac{R_{eq} - R_{pol}}{R_{eq}},
	\label{flattening}
\end{equation}

\noindent where $R_{eq}$ and $R_{pol}$ represent the equatorial and polar radii of the planet respectively. 

Knowledge of the oblateness of a planet can provide information about its rotation rate (or rotation period) and internal density structure. These can in turn shed valuable insight into the planet's formation and evolution \citep{lissauer, li_lai}, as well as its atmospheric circulation and dynamics \citep{Kaspi_2015}.
The solar system planets have different rotation periods and oblateness indicating diverse formation and evolutionary histories \citep{laskar_robutel}. Saturn, having one of the fastest rotation with a period of only 10.7\,hrs, has the highest oblateness of $f=0.098$ (i.e., its polar radius is 9.8\% smaller than its equatorial radius). Although Saturn rotates slightly slower than Jupiter, it has a significantly lower density, and thus less gravity, which allows its rotation induce a higher oblateness than in Jupiter (with $f=0.065$).

Measuring the oblateness of exoplanets is challenging as the induced effects in transit signals have low amplitudes. Previous studies investigated the photometric difference between the transit light curve of an oblate planet and the corresponding spherical planet (e.g \citealt{seagerhui,barnes03,carterwinna}). They showed that the amplitude of the oblateness-induced signal for a  giant planet, with Saturn-like oblateness and planet-to-star radius ratio of 0.1, is just around 100\,ppm for the best case transit geometry. \citet{zhu} searched for oblateness signals in Kepler light curve data and obtained a tentatively high oblateness of 0.22 for the brown dwarf Kepler-39b  although they could not validate the consistency of the measurement across different subsets of the data. Later work by \citet{biersteker} did not detect the oblateness of Kepler-39b but put loose constraints on the oblateness of Kepler-427b. Therefore, measuring the oblateness of planets remains challenging. For very precise transit signals, assuming sphericity for an oblate planet would lead to systematic errors in the determination of the transit parameters \citep{barnes03}.

In this paper, we complement the previous works by investigating, for the first time, the signature of planet oblateness in transit spectroscopy using the Rossiter-McLaughlin (RM) effect - the radial velocity (RV) anomaly caused by a companion transiting a rotating star \citep{rossiter,mclaughlin}. Besides the usual planetary transit parameters, the RV anomaly caused by a planet additionally depends on the projected stellar rotation velocity, $v\sin{i_{\ast}}$, and the projected spin-orbit angle, $\lambda$, between the sky projections of the stellar spin axis and the planetary orbit normal \citep{gaudi_winn}. We compare the detectability of oblateness from spectroscopic RM signals to that from photometric light curves and also the prospects of combining both measurements for a more precise measurement of oblateness. 

In Sect.~\ref{sect2}, we describe the oblate planet model and transit tool used in generating oblate planet light curves and RM signals. We used the tool to confirm photometric results from previous studies while showing also the expected spectroscopic result. In Sect.~\ref{sect3}, we identify the optimal parameter space for detecting the photometric and spectroscopic oblateness signal and also investigate the detectability of oblateness considering different observing instruments. In Sect.~\ref{sect4}, we discuss the results and conclude in Sect.~\ref{sect5}. 

\section{Oblate planet transit}
\label{sect2}
As mentioned in \citet{barnes} and \citet{akin18}, the projected shape of a planet with a continuous opaque ring extending directly from the planet surface (so that there is no gap between them) will mimic the projected shape of an oblate planet; if the ring is appropriately inclined with respect to the sky plane. As such, a ringed planet model can be used to simulate the expected  photometric light curve and spectroscopic RM signal of an oblate planet. Therefore, we employ the SOAP3.0 ringed planet tool \citep{akin18} to model the transit of an oblate planet\footnote{Although there might be some slight differences in using a ringed planet model to emulate the projected shape of oblate planets, it serves as a sufficient approximation and produces desired results consistent with previous studies.}. The ring is defined by four parameters: the inner and outer ring radii,  $R_{in}$ and $R_{out}$, in units of a core planetary radius $R_{pc}$, and also two orientation angles $i_{r}$ and $\theta$  which define the inclination of ring plane with respect to the sky plane and the obliquity of ring from the orbital plane, respectively (see Fig. 1 in \citealt{akin18}). 

\begin{figure}
	\includegraphics[width=\columnwidth]{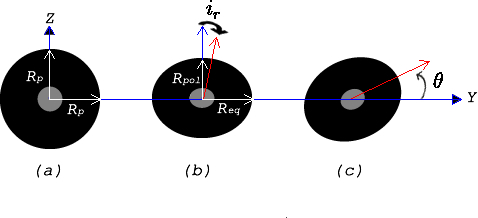}
   \caption{Schematic of the ring planet model used to describe an oblate planet. The small core planet is represented in gray and the ring in black, both of which are opaque. (a) shows the projection of a spherical planet with total radius $R_{p}$ modeled with circular face-on ring around a smaller ($0.1R_{p}$) core planet. (b) shows the ring now inclined away from sky plane YZ by an angle $i_{r}=36.87\degree$. This models an oblate planet with exaggerated $f=0.2$ making $R_{pol} < R_{eq}$. (c) shows the oblate planet now tilted from the orbital axis Y by an obliquity angle $\theta=+30\degree$.}
    \label{oblate}
\end{figure}

As depicted in Fig.~\ref{oblate}, the projected shape of an oblate planet can be obtained using the ring planet model by first setting a core planet with a  negligibly small radius (e.g. 10\% of the required $R_{eq}$;  this is necessary since the inner and outer ring radii are in units of a core planet). A circular opaque ring starting at the surface of the core planet is then added with outer radius $R_{out}$ extending out to the equatorial radius of the oblate planet to be modeled. Oblateness of the entire projected figure  (core planet + ring) can be obtained by inclining the ring away from sky plane by $i_{r}=\cos^{-1}(1-f)$  which imitates a reduced radius at the poles compared to the equator. As $i_{r}$ increases, the total projected figure becomes more oblate ($f$ increases). The obliquity of the  ring $\theta$ also corresponds to the obliquity of the oblate planet defining the projected angle between its equatorial plane and the orbital plane. It ranges from $-90\degree$ to $+90\degree$ with positive angles measured anti-clockwise from the transit chord and negative angles measured clockwise. 

The rotation period ($P_{rot}$) of the planet is related to the oblateness ($f$) by
\begin{equation}
       P_{rot} = 2\pi\sqrt{\frac{R_{eq}^3}{GM_{p}(2f-3J_{2})}}~,
	\label{rotperiod}
\end{equation}

\noindent where $J_{2}$ and $M_{p}$ represent the quadrupole moment and  mass of the planet,  respectively \citep{carterwinna}. This equation shows that $f$ is inversely proportional to the rotation period and density ($\rho_p \approx M_p/R_{eq}^3$) of the planet, so the effect of oblateness will be most significant for gaseous planets with short rotation periods.

We point out that $f$ and $\theta$ measured from transit light curves are not the true planet oblateness and obliquity but their projection on the sky plane since they are derived from the projected shape of the transiting planet (an ellipse). This means that the transit-derived values of  $f$ and $\theta$ will always be lower limits on the true values. This, in turn, implies that only an upper limit on $P_{rot} $ can be obtained \citep{seagerhui,carterwinna}.

We define the mean radius of an oblate planet by
\begin{equation}
\label{mean_rad}
\bar{R}_{p}~=~\sqrt{R_{eq}R_{pol}}~=~R_{eq}\sqrt{1-f}.     
\end{equation}
So for $f=0$, $R_{eq}=R_{pol}$ and the mean radius of the oblate planet is same as the radius of a spherical planet. The maximum possible value of $f$ that can be attained by a planet is at the rotational break-up limit when the centrifugal acceleration balances the gravitational acceleration at the equator and this is at $f \simeq 0.5$ \citep{carterwinna}.

\subsection{Oblateness-induced signal in light curves and RM signals}
\label{} 
Previous studies compared the transit light curve of an oblate planet to that of a spherical planet and showed that the oblateness signal manifests itself at the ingress and egress phases (e.g., \citealt{carterwinna,zhu}). The oblateness-induced signal is a geometrical effect and is obtained as the residuals from fitting the transit observation of an oblate planet with a spherical planet model. The signal arises mostly due to difference in contact times at the stellar limb (ingress and egress) between the transiting oblate planet and the corresponding spherical planet of the same area. Since the geometry of a transiting planet is the same when taking photometric and spectroscopic transit measurements, the oblateness signature should also manifest itself in the transit RM signal.

\begin{figure}
	\includegraphics[width=\columnwidth]{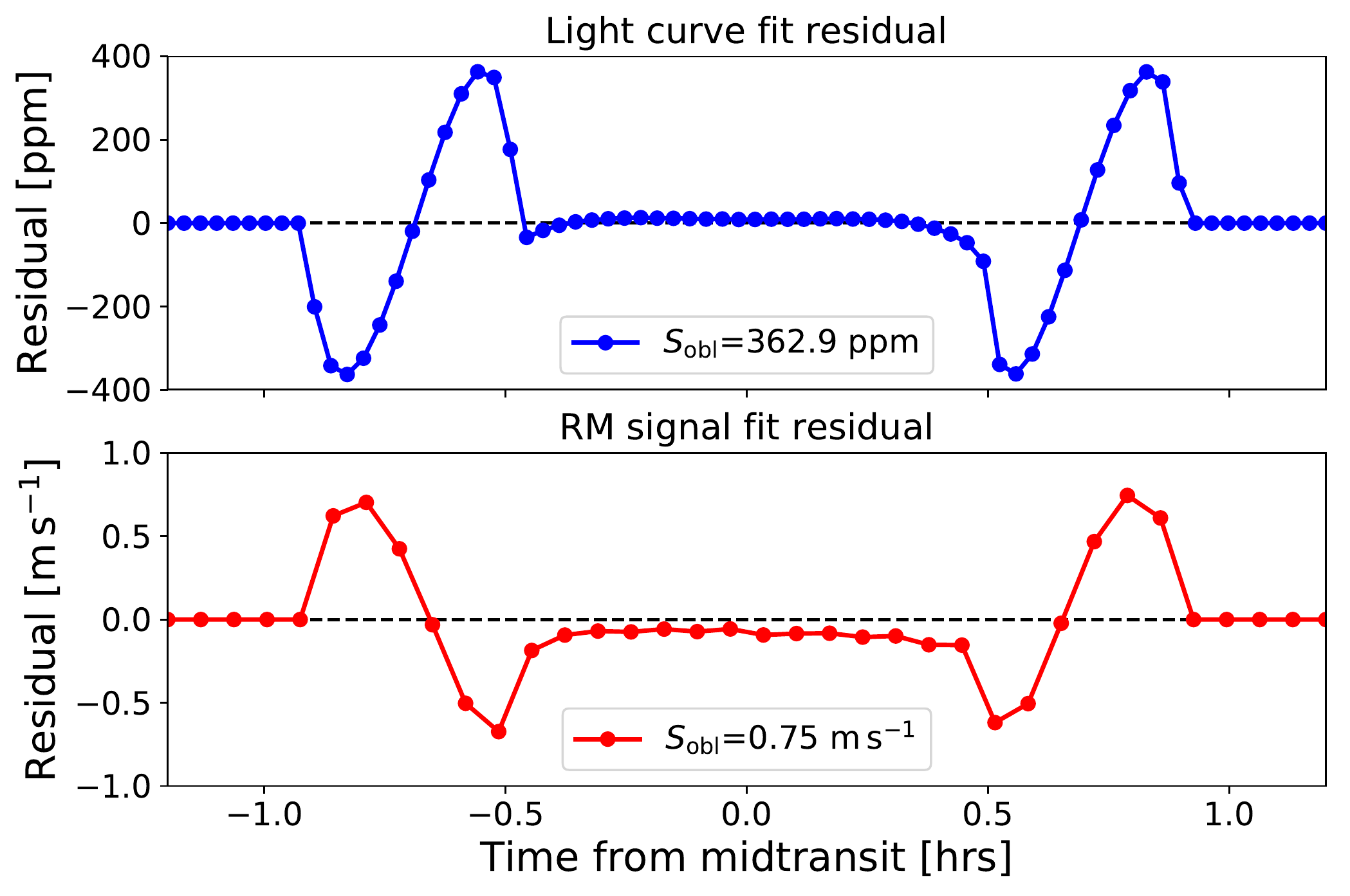}
	\caption{Oblateness-induced signatures obtained as the residuals from fitting the simulated oblate planet transit signals (top: light curve, bottom: RM signal) of HD 189733b assuming projected quantities $f=0.098$ and $\theta=45\degree$.}
    \label{obl_sig}
\end{figure}

To illustrate the oblateness signature in the light curve and RM signal,  we follow the work of \citet{carterwinna} which analysed seven \textit{Spitzer} transit observations of the giant planet HD\,189733b to constrain its oblateness. The planet was selected due to its large size, bright host star and availability of precise \textit{Spitzer} data. First, they calculated the theoretical oblateness signal amplitude expected for HD\,189733b by simulating its photometric light curve assuming Saturn-like oblateness of $f=0.098$ and projected obliquity $\theta=45\degree$. The planet has the following parameters: planet-star mean radius ratio $\Bar{R}_{p}=0.15463\,R_{\ast}$ (equivalent to $1.13\,R_{\mathrm{Jup}}$), scaled semi-major axis $a/R_{\ast}=8.81$, inclination $i_{p}=85.58\degree$ (or impact parameter $b=0.68$) and orbital period $P_{\mathrm{orb}}=2.22$\,days \citep{torres}. The host star is of K2 spectral type with V-magnitude $m_{V}$ = 7.7, radius $R_{\ast}=0.75\,R_{\odot}$ and $v\sin{i_{\ast}}=3.5$\,km\,s$^{-1}$ but we assume a  stellar inclination $i_{\ast}=90\degree$ (rotation axis parallel to sky plane).  We also simulated mock oblate planet transit signals (light curve and RM signal) for this planet using SOAP3.0. The light curve was generated with \textit{Spitzer} cadence of 2 minutes using quadratic limb darkening coefficients (LDCs) of [0.076, 0.034] in the \textit{Spitzer} infrared band as given in \citet{carterwinna}. The RM signal was generated with an integration time of 4\,mins\footnote{ This integration time is similar to the 5\,mins exposures of \texttt{HARPS} archival RM observations of HD\,189733b used by \citealt{triaud09,wyttenbach}.} which, for real observations, will help reduce the amplitude of stellar noise in this star as recommended in \citet{chaplin} for K-dwarfs. We obtained LDCs for the star in the visible band using \citet{claret} and assumed $\lambda=0\degree$ (close to the value of $0.85\degree$ derived in \citealt{triaud09}). We then separately fit the oblate planet light curve and also its RM signal with spherical planet models. The residuals (oblate - spherical) from both fits are shown in Figure \ref{obl_sig} with their amplitudes ($S_{\mathrm{obl}}$) quoted. The residuals show the effects of oblateness at ingress and egress as described in \citet{seagerhui} and \citet{barnes03}.

The residuals from our light curve fit is in agreement, in shape and amplitude, with the photometric result obtained in \citet{carterwinna} for this planet (see their Figure 1). The oblateness-induced signature in the RM signal, shown here for the first time, has an amplitude  $S_{\mathrm{obl}}=0.75$\,m\,s$^{-1}$ whereas the \texttt{ESPRESSO} spectrograph (installed on the 8m class telescopes of the ESO VLT; \citealt{pepe}) is capable of attaining instrumental RV precisions up to 0.1\,m\,s$^{-1}$. Using the \texttt{ESPRESSO} Exposure Time Calculator (ETC)\footnote{\href{https://www.eso.org/observing/etc/bin/gen/form?INS.NAME=ESPRESSO+INS.MODE=spectro}{www.eso.org/observing/etc} - Version P105.6}, we estimated that the theoretical \texttt{ESPRESSO} RV precision for this star for the 4-minute exposure time is 0.31\,m\,s$^{-1}$ for observation in the high resolution (UT1; 140,000) mode\footnote{Note: In simulating the aforementioned RM signal, we used this \texttt{ESPRESSO} resolution to calculate the FWHM of the CCF which was additionally broadened to account for macro-turbulence (convection) using calibration equation from \citet{doyle14}.}. Taking into account the rotational velocity broadening which degrades the quality factor of the spectra with respect to its non-rotating counterpart, the RV precision for this star degrades by a factor of 1.8 to $\sim$0.56\,m\,s$^{-1}$ (calculated by interpolating between values in Table 2 of \citealt{bouchy}). This precision can still be better  if data is phase folded over an increased number of transit observations (n) as it scales in the photon noise limit with $1/\sqrt{n}$.  If we assume that seven transits are observed (number of \textit{Spitzer} transits analysed in \citealt{carterwinna}), a better RV precision of $\sim$0.2\,m\,s$^{-1}$ would be achieved which is much lower than the spectroscopic oblateness amplitude of this system. Moreover, RM measurements of longer period planets can be obtained with longer exposure times that provide a higher RV precision such that several transit observations are not necessarily needed.

Although the analysis of this system is simply illustrative, it hints that \texttt{ESPRESSO} RM measurements of "relevant" systems with transiting planets could allow reasonable constraints to be placed on planet oblateness in addition to those from light curve analysis. We further investigate this possibility in the following sections with suitable considerations.  

We point out that the oblateness signal can be similar to the signal caused by the presence of rings  \citep{seagerhui,barnes}. However, depending on ring orientation and size of the gap between the planet and ring, the ring signature can be distinct showing more peaks at ingress and egress than the oblateness signature (see \citealt{ohta, akin18}). Additionally, an estimate of the planetary density can help distinguish them since a planet with rings will appear larger and cause the planetary density to be underestimated \citep{zuluaga,akin20}.

\subsection{Tidal interaction}
\label{tides}

In the previous section we showed the signature of oblateness in the photometric light curve and spectroscopic RM signal for a short period planet. However, short period planets such as HD\,189733b have short tidal evolution timescales so they are expected to already have circularized orbits and synchronised rotations such that $P_{rot}$ becomes equal to $P_{\mathrm{orb}}$ \citep{guillot}. Such short period planets cannot have significant rotation induced oblateness since their rotations will be too slow \citep{seagerhui}. Indeed, \citet{carterwinna} did not detect oblateness in this planet for this same reason inspite of the large theoretically expected photometric signal (top panel of Fig.\,\ref{obl_sig}). However, short period planets can still be radially deformed (in direction of the star) due to the strong stellar tidal interaction. Previous studies (e.g. \citealt{correia14,akin19,hellard19}) discussed the detectability of tidally induced deformation in short period planets and showed that this deformation diminishes with distance to the star.

Strong tidal interaction also affects the obliquity of short period planets by driving the value of $\theta$ to zero at the same short timescale for attaining rotation synchronization \citep{peale}. Avoiding stellar tidal interaction then imposes the requirement for long period orbits to ensure rapid planet rotation for significant oblateness. \citet{seagerhui} and \citet{carterwinnb} showed that a Jupiter-mass planet orbiting a Solar twin star at a distance of $\sim$0.2\,AU ($P_{\mathrm{orb}} \simeq 30$\,days) will have tidal synchronization timescale of $\sim$10\,Gyrs (which is greater than the age of most host stars) so that the planet is not expected to have spun down and can therefore have significant rotation.
\subsection{Spin precession}

The spin axis of an oblate planet will precess due to the gravitational torque exerted on the planet by the host star. The effect of spin precession on transit signals is that the orientation of the oblate planet changes with time and so does its projection. This leads to transit depth variations between obtained light curves and amplitude variations between the RM signals. Such variations can complicate efforts to measure the ingress and egress oblateness signature since combining successive transits might average out the subtle signal. However, a Jupiter or Saturn-like planet with $P_{\mathrm{orb}} \simeq 30$\,days is expected to have precession period of $\sim$50\,yrs \citep{carterwinnb} which is too long to be observed within a few transit observations of the planet. For example, the spin axis of a $P_{\mathrm{orb}}=30$\,days planet will precess by a negligible $1.8\degree$ in 3 successive transit observations thereby allowing the phase folding of data to probe oblateness. 

Although some studies (e.g. \citealt{carterwinnb,biersteker}) have illustrated the possibility of detecting oblateness due to spin precession for $P_{\mathrm{orb}}<30$\,days, we investigate here the case of planets with $P_{\mathrm{orb}}\geq30$\,days for which significant rotation induced oblateness is expected and spin precession is negligible.

\begin{table}
\centering
\caption{Adopted system parameters of the hypothetical long period oblate planet.}

\label{tab_parameters}

\begin{tabular}{ll}
\hline\hline
Parameter {[}Unit{]}    \qquad \qquad \qquad \qquad     & Value          \\
\hline
$\bar{R}_{p} \,[R_{\ast}]$          & 0.1546         \\
[3pt]
$a/R_{\ast}$                & 70.75        \\
[3pt]
$P_{\mathrm{\mathrm{orb}}}$ [days]             & 50.0           \\
[3pt]
$\lambda\,[\degree]$      & 0.0             \\
[3pt]
$f$                          & 0.0980         \\
[3pt]
Stellar type               & K2V       \\
$m_{V}$                      & 7.7 \\
$q_{1}$, $q_{2}$               & 0.5151, 0.3872 \\
[3pt]
$v\sin{i_{\ast}}$ {[}km\,s$^{-1}${]}  & 5.0    \\
\hline
\end{tabular}%

\end{table}

\section{Detecting oblateness}
\label{sect3}
\subsection{Identifying optimal transit geometry for detecting oblateness }
\label{favor}

As the oblateness signal can depend strongly on the projected orientation of the planet ($\theta$) and that of its orbit ($i_{p}$ or $b$), it is important to identify the combinations of these two parameters that are optimal for detecting oblateness. Even though the value of $\theta$ is unknown a-priori for planets, an understanding of how it affects the detectability of oblateness is crucial as it can be used to calculate the maximum theoretical oblateness signal expected for a planet and thus aid in target selection. Although previous works (e.g., \citealt{barnes03,zhu}) have identified $\theta-b$ parameter combinations with the highest photometric oblateness signal amplitude, we are particularly interested in understanding how the spectroscopic signal amplitude varies in comparison to the precision of new spectrographs like \texttt{ESPRESSO}. 

We take case study of the HD189733 system assuming its Jupiter-sized planet ($1.13\,R_{\mathrm{Jup}}$) has a circular orbit around the star with a longer period of 50\,days (based on requirements from previous section). As in the previous section, we assume a projected planet oblateness of $f=0.098$. We also assume that the star rotates slightly faster with $v\sin{i_{\ast}}= 5$\,km\,s$^{-1}$ (taking $i_{\ast}=90\degree$). We use quadratic LDCs for this star in the visual band obtained from \citet{claret} but re-parameterized as [$q_{1}$, $q_{2}$] following \citet{kipping13}. The full adopted parameters for the simulated system is given in Table \ref{tab_parameters}.

We used SOAP3.0 to generate theoretical transit signals (light curves and RM signals) for the adopted oblate planet on a grid of obliquity ($\theta$) and impact parameter ($b$) values with a total of 42 combinations (excluding grazing transits since their ingress/egress are undefined).  The transit durations and ingress (egress) durations of the transit signals vary with $b$. From  $b$\,=\,0 to $b$\,=\,0.8, the transit durations decrease from $\sim$6.3\,hrs to $\sim$4.5\,hrs whereas the ingress (egress) durations increase  from $\sim$55\,mins to $\sim$90\,mins. Transits at higher impact parameters will thus allow better sampling of oblateness induced features due to their longer ingress and egress durations. The light curves were simulated with 2-min integration time  (or binning) similar to \texttt{TESS} but applicable to other photometric instruments to increase the attained precision of each measurement. The RM signals had a longer integration time of 8\,mins\footnote{ Spectroscopic transit observations usually require longer integrations than the photometric because spectrographs lose photons due to slit losses, stray light, and scattered light and so require longer time to collect more photons \citep{oshagh17}. Long integrations are also used to average out and reduce stellar RV noise. We note also that longer period planets can have longer integrations which should be chosen such that the ingress and egress are still well sampled.} which enables \texttt{ESPRESSO} reach  a higher RV precision of 0.22\,m\,s$^{-1}$ while still providing good temporal resolution of the ingress and egress of this planet. Other spectrographs (like \texttt{HARPS} installed on a smaller 3.6\,m telescope) are not capable of reaching such precision within this short integration time.  To attain the same precision as \texttt{ESPRESSO}, \texttt{HARPS} will require $\sim$30\,mins integrations which will not allow enough data points to probe oblateness signatures at ingress and egress. 

We perform a least squares fit to the oblate planet light curve and RM signal of each parameter combination using a spherical planet model and obtain the amplitude, $S_{\mathrm{obl}}$, of the residuals (at ingress/egress).  The free parameters in the light curve fit were $R_{p}$, $a/R_{\ast}$, $b$, $q_{1}$, and, $q_{2}$ while the RM signal fit additionally had $v\sin{i_{\ast}}$ and $\lambda$.

\begin{figure}
	\includegraphics[width=\columnwidth]{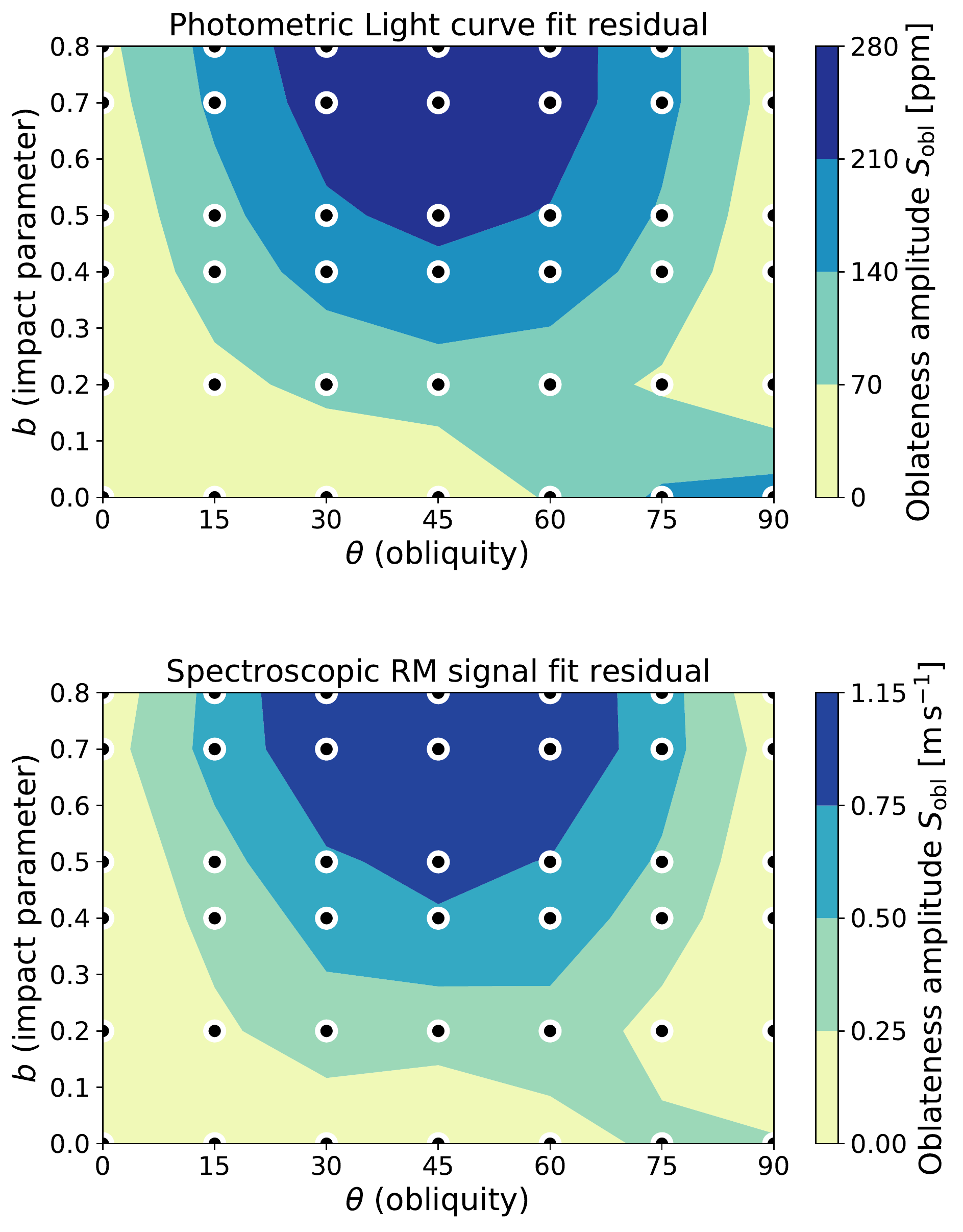}
	\caption{Contour plots showing amplitude, $S_{\mathrm{obl}}$, of observable oblateness-induced signal. The plots are generated from fitting 42 oblate planet transit signals of different $\theta - b$ combinations with a spherical planet model. Top: Contour of light curve fit residuals. Bottom: Contour of RM signal fit residuals. Black circles indicate the parameter combinations of the signals from which the residual amplitudes were obtained.}
    \label{contour}
\end{figure}

Figure \ref{contour} shows the contour plots generated using the residual amplitudes from fitting the  photometric light curve and spectroscopic RM signal of each parameter combination. It represents the amplitude of observable oblateness-induced signal, $S_{\mathrm{obl}}$, at each parameter combination.  The contours are shown only for positive $\theta$ angles as the pattern is symmetric about $\theta=0$. In both contour plots, we see a similar trend showing that the amplitude of  photometric and spectroscopic oblateness signal is lowest  (yellow regions) at zero obliquity across all impact parameters. This implies that the best-fit spherical model can easily emulate the light curve and RM signal of the oblate planet by adjusting its fit parameters thereby making it difficult to detect oblateness at these orientations. However, at higher obliquities, the amplitude increases with impact parameter due to asymmetry between ingress and egress of the oblate planet transit signal that cannot be easily emulated by a spherical model. The oblateness signal reaches its peak amplitude at points around $\theta=45\degree$, $b=0.7$  (dark blue regions) in agreement with the results of earlier  photometric studies (e.g. \citealt{seagerhui,carterwinna}) but confirmed here to be same in  spectroscopy. The contour plot is nearly symmetric about the vertical $45\degree$ line with the exception of a few orientations around $b,\,\theta=0,\,90\degree$. 

 We also investigated how the spin-orbit angle ($\lambda$) affects the amplitude of oblateness signal. For the yellow regions in Fig. \ref{contour} with low oblateness signal amplitudes (e.g. orientations with $\theta\approx0\degree$ and those with $b\approx0$), we find that the  spectroscopic oblateness signal amplitudes can increase for planets with  spin-orbit misalignment (i.e. $\lambda$\,$\neq$\,0). For example, the  spectroscopic oblateness amplitude at the orientation $\theta$\,=\,$0\degree$,\,$b$\,=\,0.2 is only 0.05\,m\,s$^{-1}$ when $\lambda=0\degree$ but significantly increases to 0.61\,m\,s$^{-1}$ for $\lambda=30\degree$. In contrast, this orientation has a photometric oblateness amplitude of only 3\,ppm irrespective of $\lambda$. The orientations in the yellow regions can thus be more favorable for detecting oblateness in  spectroscopy than in photometry. Figure \ref{lbdafits} shows the variation of the spectroscopic oblateness amplitude with $\lambda$ at orientations with $\theta=0\degree$ and also orientations with $b=0$.

\begin{figure}
	\includegraphics[width=\columnwidth]{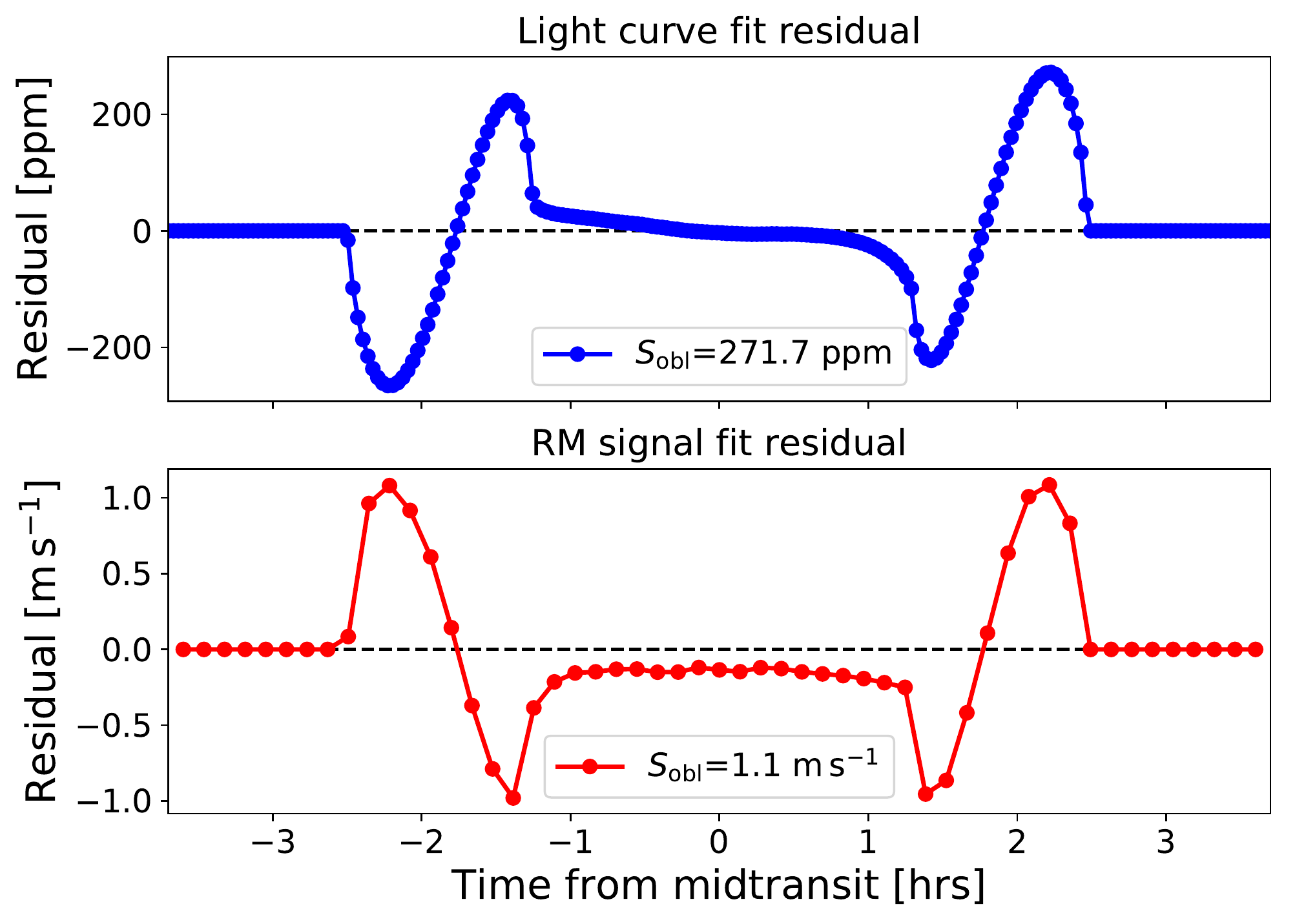}
	\caption{Oblateness-induced signal obtained from fitting simulated oblate planet of $\theta=45\degree$, $b=0.7$ with spherical planet model. Top: light curve fit residual. Bottom: RM signal fit residual. }
    \label{45_07}
\end{figure}

In summary, Fig. \ref{contour} shows that the oblateness-induced signal is more prominent,  in photometry and spectroscopy, for oblique planets at high impact parameters particularly for the points around $\theta=45\degree$, $b=0.7$ (optimal transit orientation) which provides the best opportunity to measure oblateness. The residual from the fit of this orientation is shown in Fig. \ref{45_07} with  photometric and spectroscopic oblateness amplitudes of $\sim$272\,ppm and 1.1\,m\,s$^{-1}$, respectively.

The amplitude of the photometric and spectroscopic oblateness-induced signal for a particular parameter combination ($\theta,\,b$) scales with the oblateness and planet radius as $\sim$\,$(f/0.098)\,(\bar{R}_{p}/0.1546)^2$ \citep{barnes03}. Therefore, planets with larger planet-to-star radius ratios will have larger oblateness signatures for the same projected oblateness. For the same planet, the spectroscopic oblateness amplitude ($S_{\mathrm{obl}}^{~~\mathrm{sp.}}$) additionally scales with the projected stellar rotational velocity following the relation
\begin{equation}
   S_{\mathrm{obl}}^{~~\mathrm{sp.}}~ \propto ~ \left(\frac{v\sin{i_{\ast}}}{5\,\mathrm{km\,s}^{-1}}\right)\, \left(\frac{f}{0.098}\right)\, \left(\frac{\bar{R}_{p}}{0.1546}\right)^2 .
\end{equation}

\subsection{Oblateness detectability}
\label{section3.2}
Detecting planet oblateness implies obtaining a measurement for the parameter $f$ by fitting a transit observation, of sufficient precision, with an oblate planet model. Recovering $f\simeq0$ implies that a spherical planet model is a better fit to the observation than an oblate planet model. Detecting oblateness thus requires that $f$ is recovered with accuracy and statistical significance above zero from the fitting process.  

To illustrate the photometric and spectroscopic detectability of oblateness, we simulated the transit signals (light curve and RM) of the oblate planet at the determined optimal orientation (Table~\ref{tab_parameters} with $\theta\,=\,45\degree,\,b\,=\,0.7$) which have oblateness amplitudes of 272\,ppm and 1.1\,m\,s$^{-1}$ (Fig.~\ref{45_07}). The light curve and RM signal were simulated with cadences of 2\,mins and 8\,mins, respectively. Random Gaussian (white) noise of different levels, $N$, was added to the simulated transit signals. We then investigated how well we can recover $f$ and at what noise level it would be difficult to distinguish between the oblate planet and spherical planet transit signals (light curve and RM). This is useful in order to understand the instrumental precisions required for photometric and spectroscopic detection of oblateness.

We performed a Markov Chain Monte Carlo (MCMC) analysis  with SOAP3.0 model to recover the oblate planet transit parameters along with their uncertainties. We use the \texttt{emcee} package \citep{foreman} with priors on the parameters as given in Table. \ref{priors}. We assume that the LDCs can be known with 10\% accuracy so we adopt gaussian priors on them centered on the true simulated values with 10\% standard deviation. The MCMC was performed with 36 walkers each having 20,000 steps (which was several times the computed auto-correlation time as recommended in \texttt{emcee} as a convergence diagnostic). The initial 25\% of the steps were then discarded as burn-in. 

\begin{figure*}
\centering
    \includegraphics[width=17.5cm]{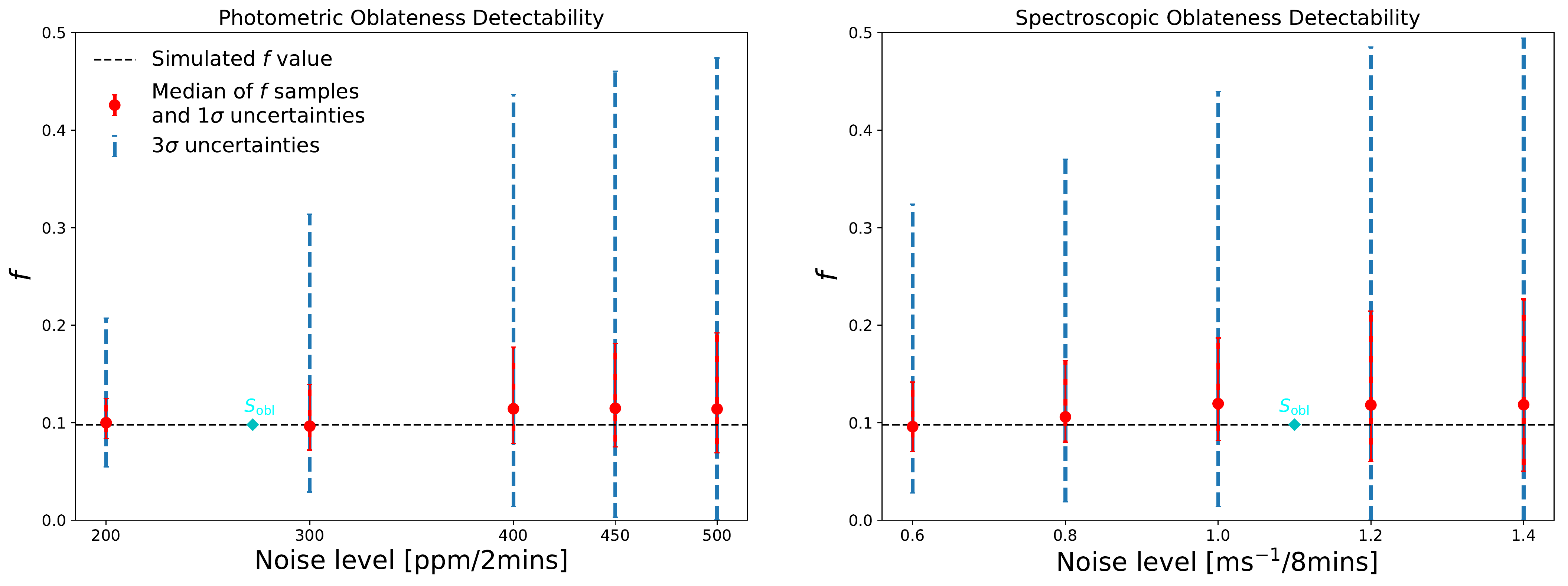}
    \caption{Detectability of oblateness in photometry (left) and RV (right) as a function of instrumental noise. The black dashed line is the simulated $f$ value. The points are the median of the $f$ samples at each noise level. The red errorbars show the 68\% credible interval ($\pm1\sigma$) while the blue errorbars show the 99.7\% credible interval ($\pm3\sigma$). The cyan diamonds, labeled $S_{\mathrm{obl}}$, indicates the photometric and RV oblateness signal amplitude as  obtained in Fig~\ref{45_07}.}
    \label{detectability}
\end{figure*}

 Figure~\ref{detectability} shows the oblateness detectability plot, in photometry and  spectroscopy, indicating the median and standard deviations of the recovered $f$ at different noise levels (average of three MCMC realizations).
  For oblateness to be confidently detected ($f$ measured), we require that $f$ is recovered with $3\sigma$ significance above zero.\footnote{ As a check of our MCMC analysis, we confirmed that fitting a spherical planet transit signal with an oblate planet model recovers $f\simeq0$ at different noise levels.} We see in the spectroscopic detectability plot (right panel of Fig.~\ref{detectability}) that $f$ is detected with $3\sigma$ significance for noise levels up to 1\,m\,s$^{-1}$/8\,mins. At higher noise levels, the distribution of the recovered $f$ samples include $f=0$ at $3\sigma$ implying that a spherical planet model is also probable. In the photometric detectability plot (left panel of Fig.~\ref{detectability}), a $3\sigma$ detection is attained for noise levels up to 400\,ppm/2\,mins. We note that $2\sigma$ detection of oblateness can still be attained at higher noise levels (up to 550\,ppm in photometry and 1.5\,m\,s$^{-1}$ in spectroscopy).

\begin{table}
\centering
\caption{Priors on the fitted parameters in the MCMC. $f$ ranges from zero for a spherical planet model to the maximum possible value of 0.5.  $~^{+}$ denotes additional prior parameters used in the RM signal MCMC.}
\label{priors}
\begin{tabular}{lll}
\hline\hline
Parameter               &\qquad  Prior         &\qquad Interval     \\
\hline
$\bar{R}_{p}$           &\qquad  Uniform      &\qquad $\mathcal{U}(0.10, 0.20)$ \\
 [2pt]
$b$                     &\qquad Uniform       &\qquad $\mathcal{U}(0, 1)$ \\
 [2pt]
$a/R_{\ast}$            &\qquad Uniform       &\qquad $\mathcal{U}(80, 120)$  \\
[2pt]
$q_{1}$                 &\qquad Gaussian      &\qquad $\mathcal{N}(0.5151, 0.052)$   \\
 [2pt]
$q_2$                   &\qquad Gaussian      &\qquad $\mathcal{N}(0.3872, 0.039)$   \\
 [2pt]
$\theta$                &\qquad Uniform       &\qquad $\mathcal{U}(-90, 90)$  \\
 [2pt]
$f$                     &\qquad Uniform     &\qquad $\mathcal{U}(0.0, 0.5)$  \\
 [2pt]
$\lambda~^{+}$              &\qquad Uniform      &\qquad $\mathcal{U}(-45, 45)$  \\
 [2pt]
$v\sin{i_{\ast}}~^{+}$       &\qquad Gaussian      &\qquad $\mathcal{N}(5.0, 0.25)$  \\

\hline
\end{tabular}
\end{table}

\begin{figure*}
	\includegraphics[width=8.812cm]{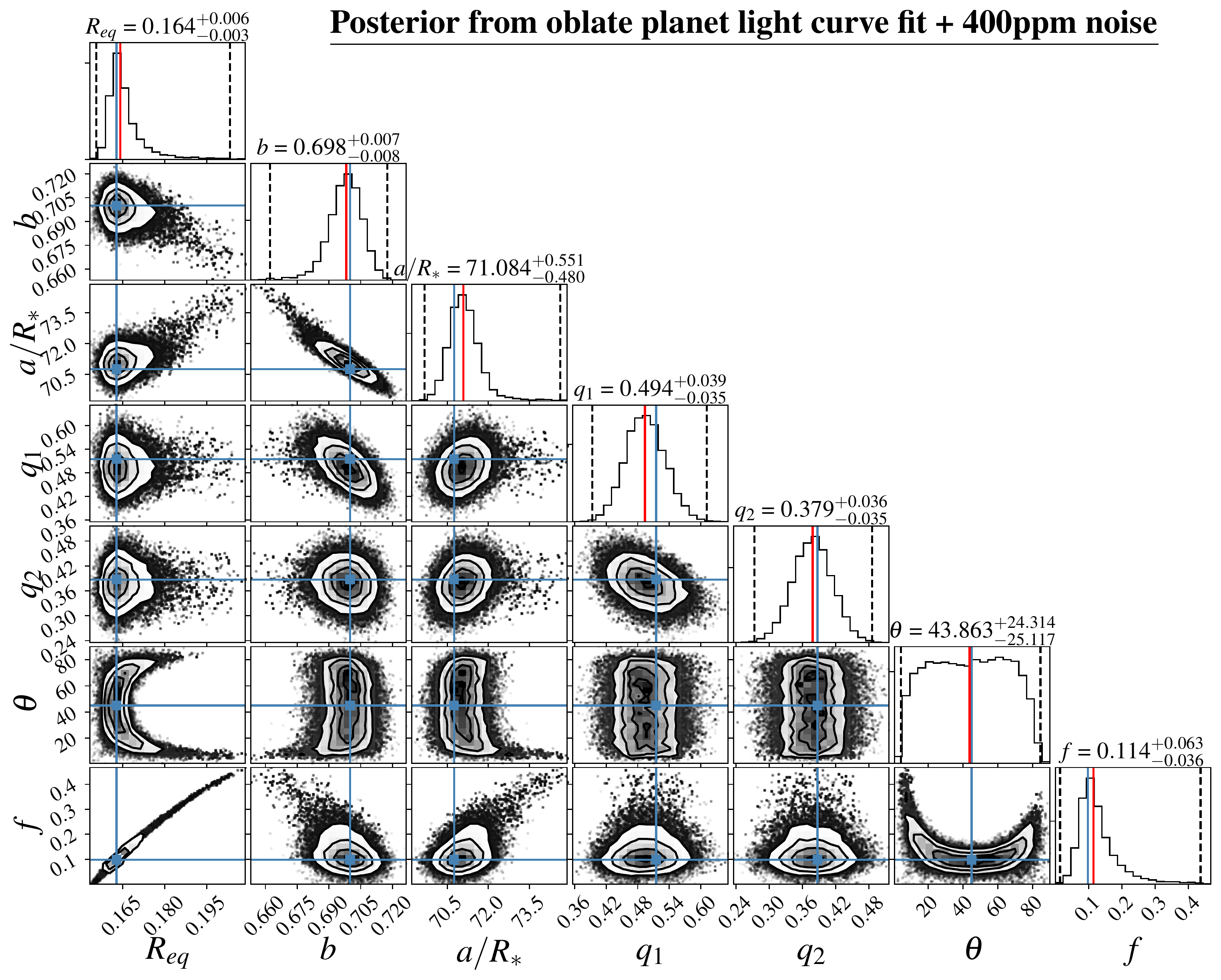}
		\includegraphics[width=8.812cm]{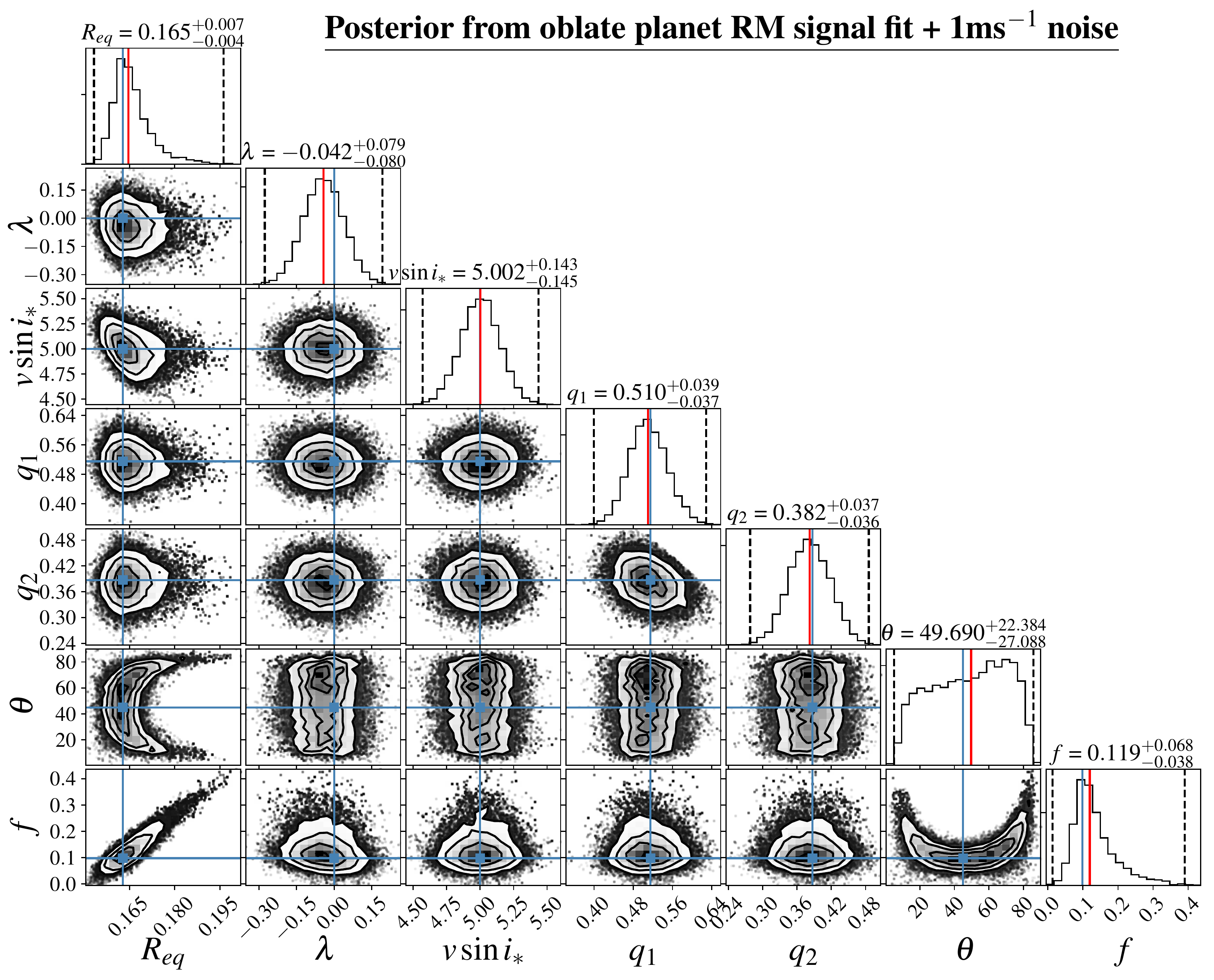}

	\caption{Left: Posterior distributions of all parameters from fitting the simulated oblate planet light curve with 400\,ppm noise added. Right: Posteriors of some parameters (distribution for $b$ and $a/R_{\ast}$ are not shown for brevity) from fitting of the simulated oblate planet RM signal with 1\,m\,s$^{-1}$ noise added. The quoted values are the median marginalized distribution of each parameter (also red lines) with the surrounding 68\% credible interval ($\pm1\sigma$). The dashed vertical lines indicate the $\pm3\sigma$ limits calculated as the $0.15^{\mathrm{th}}$ and $99.85^{\mathrm{th}}$ percentiles. Blue lines indicate the true simulated values.}
    \label{posterior}
\end{figure*}

Figure \ref{posterior} shows realized posterior distributions of the retrieved parameters at the detectable noise limits (400\,ppm noise added to light curve and 1\,m\,s$^{-1}$ added to RM signal). In both cases, we see from the $f-\theta$ joint distributions that $\theta$ is not well-constrained and is strongly correlated with $f$ ($\theta$ is  only better constrained for highly precise transit signals). Figure~\ref{contour} already showed that the amplitude of the oblateness signal is fairly symmetric about $\theta=45\degree$ and reduces as the value of $\theta$ gets farther from $45\degree$. Therefore, for $\theta$ values different from  $45\degree$ in the $f-\theta$ distribution, the amplitude of oblateness signal reduces and a higher value of $f$ is needed to fit the observation. The degeneracy between $f$ and $\theta$ is responsible for the long tail towards large oblateness in the $f$ distribution (also seen for all noise levels in Fig.~\ref{detectability}).

We converted the posterior of the fitted mean radius $\bar{R}_{p}$ to $R_{eq}$ using equation \ref{mean_rad} to show the evidently strong correlation between $f$ and $R_{eq}$; as the oblateness increases, the equatorial radius gets more elongated compared to the polar radius. It is interesting to see that the limb darkening parameters ($q_1$, $q_2$) are not strongly correlated with $f$ implying that very precise determination of the LDCs are not required to detect planetary oblateness  which is contrary to the case for detecting rings \citep{akin18}. Indeed when we adopt a non-informative uniform prior of $\mathcal{U}(0, 1)$ on the LDCs, $f$ is still similarly recovered but with larger uncertainties on the LDCs and on the other parameters. The posteriors from the RM signal MCMC (left panel in Fig.~\ref{posterior}) also shows that $\lambda$ and $v\sin{i_{\ast}}$ are not correlated with $f$ implying that planetary oblateness does not affect the accurate retrieval of these parameters from RM signals.

\begin{figure}
\centering
	\includegraphics[width=0.75\columnwidth]{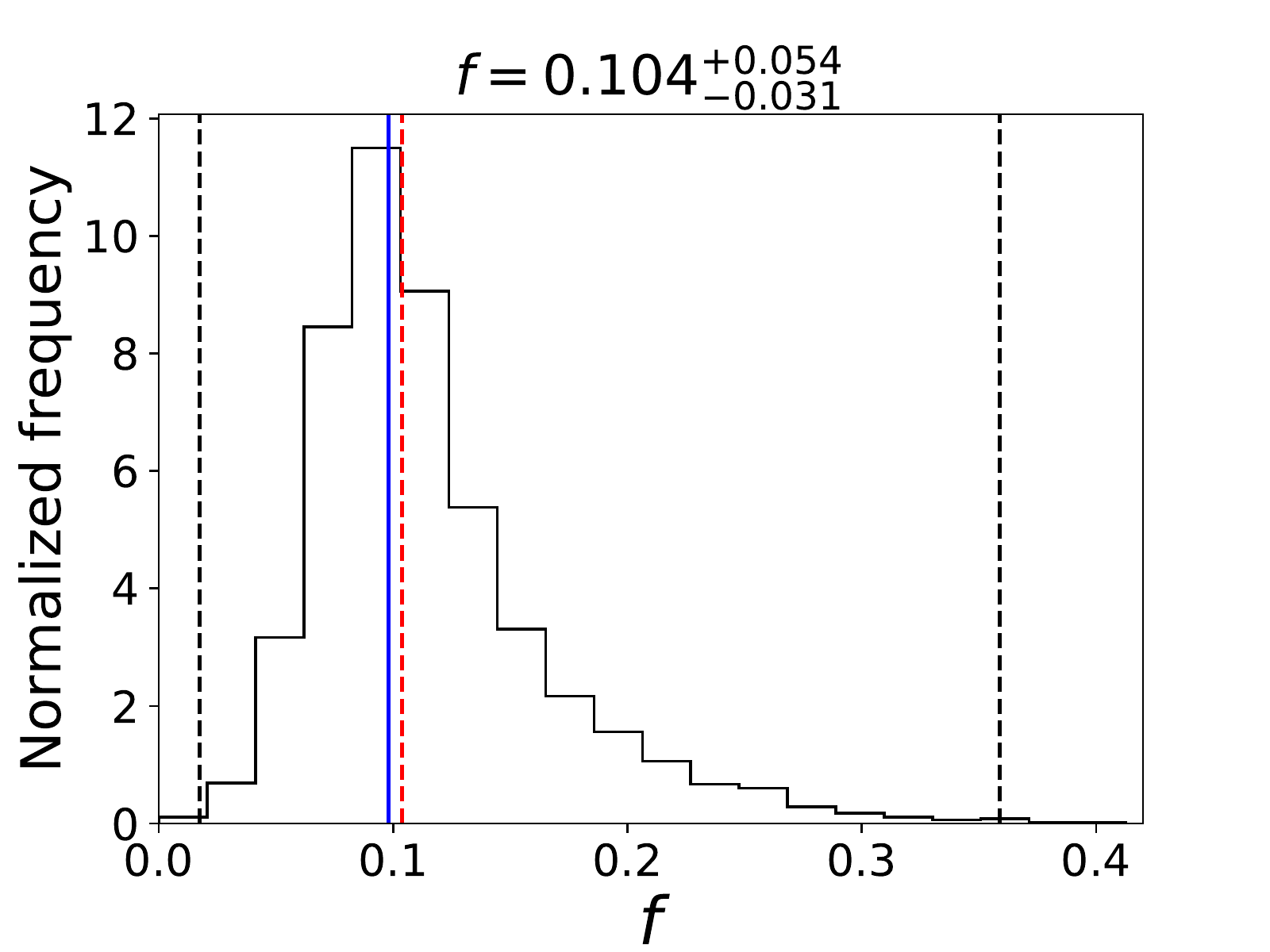}
    \caption{Histogram of recovered $f$ samples from simultaneous MCMC fit of light curve (with 450\,ppm noise) and RM signal (with 1.2\,m\,s$^{-1}$ noise). Lines and values are as described in Fig.~\ref{posterior}.}
   \label{simultaneous}
\end{figure}

We investigated also the retrieval of $f$ from combined analysis of the photometric light curve and spectroscopic RM signal. As seen in Fig.~\ref{detectability}, $f$ is not recovered with $3\sigma$ for  noise levels of 450\,ppm (added to the light curve) and 1.2\,m\,s$^{-1}$ (added to the RM signal). However, simultaneously fitting both transit signals allows the recovery of $f$ with $3\sigma$ significance and with better accuracy as shown in Fig.~\ref{simultaneous} (compared to the Figs.~\ref{detectability} and \ref{posterior}).

\begin{table}

\caption{Photon noise level per integration/binning time for stars of $m_{V}$=8 and 10 observed with different instruments}
\label{noise}
\begin{tabular}{lll}
\hline\hline
Instrument              &  Noise/time          &  Noise/time       \\
                        &  @$m_{V}=8$          &  @$m_{V}=10$       \\
\hline                                          
TESS\,$^{\mathrm{a}}$           &  512\,ppm/2\,mins         &  1109\,ppm/2\,mins     \\
[3pt]
CHEOPS\,$^{\mathrm{b}}$        &  127\,ppm/2\,mins      &  219\,ppm/2\,mins      \\
 [3pt]
PLATO\,$^{\mathrm{c}}$         &  44\,ppm/2\,mins       &  148\,ppm/2\,mins      \\
 [3pt]
JWST (NIRCam)\,$^{\mathrm{d}}$ &  $\sim$40\,ppm/2\,min       &     72\,ppm/2\,mins   \\
 [5pt]
ESPRESSO\,$^{\mathrm{e}}$      & 0.25\,m\,s$^{-1}$/8\,mins &  0.65\,m\,s$^{-1}$/8\,mins \\
[1pt]
(@$v\sin{i_{\ast}}$=5\,km\,s$^{-1}$)$^{\mathrm{\ast}}$    & (0.58\,m\,s$^{-1}$/8\,mins )   & (1.5\,m\,s$^{-1}$/8\,mins)  \\
\hline
\end{tabular}
\footnotesize
$^{\mathrm{a}}$ \href{https://heasarc.gsfc.nasa.gov/cgi-bin/tess/webtess/wtv.py}{heasarc.gsfc.nasa.gov/cgi-bin/tess/webtess/wtv.py}\\
$^{\mathrm{b}}$  \href{https://cheops.unige.ch/pht2/exposure-time-calculator/}{cheops.unige.ch/pht2/exposure-time-calculator}\\
$^{\mathrm{c}}$ \citet{rauer}\\
$^{\mathrm{d}}$\,Noise floor from \citet{beichman}. Also noise estimate from \citep{hellard19}.\\
$^{\mathrm{e}}$ \href{https://www.eso.org/observing/etc/bin/gen/form?INS.NAME=ESPRESSO+INS.MODE=spectro}{www.eso.org/observing/etc}. $^{\mathrm{\ast}}$  The RV noise level increases by a factor of 2.3 for $v\sin{i_{\ast}}$\,=\,5\,km\,s$^{-1}$ \citep{bouchy}.
\end{table}

\section{DISCUSSION}
\label{sect4}
 
As shown in Fig.~\ref{detectability}, white noise levels of 400\,ppm in photometry and 1\,m\,s$^{-1}$ in spectroscopy are the detectability noise limits required to measure Saturn-like oblateness in our hypothetical HD189733b-like planet (Table~\ref{tab_parameters}). We compare these detectability limits with the noise level attainable by different observing instruments for stars of different magnitudes as given in Table~\ref{noise}. The theoretical \texttt{ESPRESSO} noise levels for $8\,m_{V}$ and $10\,m_{V}$ stars after considering the effect of stellar rotation are 0.58\,m\,s$^{-1}$  and 1.5\,m\,s$^{-1}$ per 8\,min integrations respectively. The spectroscopic detectability noise limit of 1\,m\,s$^{-1}$ is only attainable for $m_{V}=8$ stars which implies that \texttt{ESPRESSO} is capable detecting the oblateness of planets mostly around bright stars ($m_{V} \leq 9$). For stars of $m_{V}=10$, \texttt{ESPRESSO} can still constrain the  oblateness but with a lesser significance ($<2\sigma$). In contrast, the photometric detectability noise level of 400\,ppm/2min is well attainable by \texttt{CHEOPS}, \texttt{PLATO} and \texttt{JWST} for $3\sigma$ oblateness detection for planets transiting stars as faint as $m_{V}=12$. \texttt{TESS} is only capable of measuring $f$ for the brightest stars ($m_{V} \leq 8$) and with $ \leq 2\sigma$ significance. 

In general, we define $S_{\mathrm{obl}}/N$ as the ratio of the expected oblateness signal amplitude to the observational noise level which can be used to set a baseline for detecting oblateness. A $3\sigma$ detection of Saturn-like oblateness for our hypothetical planet thus requires $S_{\mathrm{obl}}/N\geq 0.68$ (i.e. 272\,ppm/400\,ppm) for photometry and $S_{\mathrm{obl}}/N\geq 1.1$ (i.e. 1.1\,m\,s$^{-1}$/1\,m\,s$^{-1}$) for spectroscopy. However, a $2\sigma$ detection only requires $S_{\mathrm{obl}}/N$ of 0.5 and 0.73 for photometry and spectroscopy, respectively. Therefore, for a planet of $\bar{R}_{p}\,=\,0.1\,R_{\ast}$ with photometric and spectroscopic $S_{\mathrm{obl}}$ of 103\,ppm and 0.42\,m\,s$^{-1}$ respectively, 
a $2\sigma$ detection of oblateness will require noise level, $N$, below 103/0.5\,=\,206\,ppm in the photometric light curve and below 0.42/0.73\,=\,0.58\,m\,s$^{-1}$ in spectroscopic RM signal. Comparing with Table~\ref{noise}, the required noise level can be attained by \texttt{CHEOPS} for $m_{V} \leq 10$, by \texttt{PLATO} and \texttt{JWST} for even fainter stars and by \texttt{ESPRESSO} for stars with $m_{V} \leq 8$. However, noise levels can be lowered to favour more significant detection if several transits are combined.

We found that the photometric data is more sensitive for detecting oblateness as it is able to recover $f$ at lower $S_{\mathrm{obl}}/N$ than in spectroscopy. This is because we are, in general, able to sample light curves with higher cadence than RM signals thereby providing more information on the ingress/egress anomaly induced by oblateness. Nonetheless, a simultaneous fit of the light curve and RM allows consistent parameter values to be derived for the system and provides the best recovery of $f$ both in precision and accuracy. Furthermore, spectroscopic detection of oblateness would provide independent and complementary verification of the oblateness signal which will increase the credibility of any detection.

We also found that the  amplitude of photometric and spectroscopic oblateness signal can increase by more than 30\%  for observations at near-infrared (NIR) wavelengths where limb darkening is less significant. In this case, the stellar limb will be almost as bright as the center thereby amplifying the difference between the oblate and spherical planet at ingress/egress. Performing the MCMC analyses with LDCs in NIR allows recovery of $f$ with almost twice the precision at visual wavelengths and also more precise determination of $\theta$. The NIR instruments (NIRCam or NIRSpec) on the forthcoming \texttt{JWST} will be able to leverage on this to detect oblateness with greater ease.  Spectrographs operating in NIR such as \texttt{NIRPS} \citep{nirps}, \texttt{CARMENES} \citep{carmenes}, and \texttt{SPIRou} \citep{spirou} can also be beneficial although, as they are installed on 4m class telescopes, they cannot achieve RV precisions as high as \texttt{ESPRESSO}.

Given that the fast  planetary rotation capable of inducing significant oblateness is expected principally from long period planets, ground-based instruments such as \texttt{ESPRESSO} will have challenges in continuously observing transits that lasts longer than a single night. At the minimum, the observations would need to cover ingress and egress phases to probe oblateness signature if the phases align with different observation nights. Globally coordinated observations alongside other high-precision spectrographs in strategic sites can help acquire better transit coverage but will require accounting for the offset between the different spectrographs.  

Although we have assumed only photon (white) noise in our analyses, we expect other noise sources to be present in the data which will impact our detectability estimates and require lower noise levels or higher number of transits. These noise sources can originate from instrumental effects like thermal instabilities and also astrophysical effects like stellar granulation and oscillation \citep{chiavassa} and stellar activity (active regions) \citep{oshagh2013, oshagh18}. These kinds of red noise in the data can likely be tackled efficiently using gaussian processes \citep{dfm17,faria19}. The effect of stellar activity is minimal in NIR thereby giving a detection advantage to instruments operating in this wavelength range.

\section{Conclusions}
\label{sect5}

Long period giant planets are capable of rapid rotations that can cause significant planetary oblateness. While previous studies have focused on probing oblateness by analysing transit light curves, we showed that the oblateness-induced signal can also be observed in high-precision RM signals. Using the test case of a hypothetical HD\,189733b-like planet with orbital period of 50\,days, we found that such a planet with Saturn-like oblateness can cause spectroscopic oblateness signatures with large enough amplitudes to be detected by high-precision spectrographs (like \texttt{ESPRESSO}). This is especially the case for planets around rapidly-rotating stars where the spectroscopic oblateness signal is amplified. We found that the photometric and spectroscopic oblateness signals are more prominent for high impact parameter transits of oblique planets and for planets with larger planet-to-star radius ratios. Additionally, we found that planet spin-orbit misalignment can cause high amplitude spectroscopic oblateness signals at some transit orientations where the photometric signals are undetectably low. This makes detecting oblateness in RM signals more favorable over light curves at these orientations.

We showed that the photometric oblateness signal can be detected at lower signal-to-noise ratios than the spectroscopic signal principally due to better temporal resolution of light curves that allows the ingress and egress oblateness signatures to be well sampled. However, combined analyses of the light curve and RM signal of a planet can increase the precision and accuracy of oblateness measurement.

\texttt{ESPRESSO} alongside photometric instruments such as \texttt{CHEOPS}, \texttt{PLATO} and \texttt{JWST} will be capable of detecting oblateness in suitable targets. However, stellar noise associated with stellar p-mode oscillation, granulation and activity will be a strong limiting factor to real photometric and spectroscopic determination of oblateness for a given star.

\section*{Acknowledgements}
 This work was supported by Funda\c c\~ao para a Ci\^encia e a Tecnologia (FCT, Portugal)/MCTES through national funds by FEDER - Fundo Europeu de Desenvolvimento Regional through COMPETE2020 - Programa Operacional Competitividade e Internacionaliza\c c\~ao by these grants: UID/FIS/ 04434/2019, PTDC/FIS-AST/28953/2017 \& POCI-01-0145-FEDER-028953, and PTDC/FIS-AST/32113/ 2017 \& POCI-01-0145-FEDER-032113. BA acknowledges support from FCT through the FCT PhD programme PD/BD/135226/2017. S.C.C.B. also acknowledges support from FCT through Investigador FCT contracts IF/01312/2014/CP1215/CT0004. MO acknowledges the support of the DFG priority program SPP 1992 "Exploring the Diversity of Extrasolar Planets (RE 1664/17-1)". MO and BA also  acknowledge support from the FCT/DAAD bilateral grant 2019 (DAAD ID: 57453096).
 
 \section*{Data availability}
 The data generated in this article will be shared on reasonable request to the corresponding author.




\bibliographystyle{mnras}
\bibliography{references} 

\begin{thebibliography}{}
\makeatletter
\relax
\def\mn@urlcharsother{\let\do\@makeother \do\$\do\&\do\#\do\^\do\_\do\%\do\~}
\def\mn@doi{\begingroup\mn@urlcharsother \@ifnextchar [ {\mn@doi@}
  {\mn@doi@[]}}
\def\mn@doi@[#1]#2{\def\@tempa{#1}\ifx\@tempa\@empty \href
  {http://dx.doi.org/#2} {doi:#2}\else \href {http://dx.doi.org/#2} {#1}\fi
  \endgroup}
\def\mn@eprint#1#2{\mn@eprint@#1:#2::\@nil}
\def\mn@eprint@arXiv#1{\href {http://arxiv.org/abs/#1} {{\tt arXiv:#1}}}
\def\mn@eprint@dblp#1{\href {http://dblp.uni-trier.de/rec/bibtex/#1.xml}
  {dblp:#1}}
\def\mn@eprint@#1:#2:#3:#4\@nil{\def\@tempa {#1}\def\@tempb {#2}\def\@tempc
  {#3}\ifx \@tempc \@empty \let \@tempc \@tempb \let \@tempb \@tempa \fi \ifx
  \@tempb \@empty \def\@tempb {arXiv}\fi \@ifundefined
  {mn@eprint@\@tempb}{\@tempb:\@tempc}{\expandafter \expandafter \csname
  mn@eprint@\@tempb\endcsname \expandafter{\@tempc}}}

\bibitem[\protect\citeauthoryear{{Akinsanmi}, {Oshagh}, {Santos}  \&
  {Barros}}{{Akinsanmi} et~al.}{2018}]{akin18}
{Akinsanmi} B.,  {Oshagh} M.,  {Santos} N.~C.,   {Barros} S.~C.~C.,  2018,
  \mn@doi [\aap] {10.1051/0004-6361/201731215}, \href
  {https://ui.adsabs.harvard.edu/#abs/2018A&A...609A..21A} {609}

\bibitem[\protect\citeauthoryear{{Akinsanmi}, {Barros}, {Santos}, {Correia},
  {Maxted}, {Bou{\'e}}  \& {Laskar}}{{Akinsanmi} et~al.}{2019}]{akin19}
{Akinsanmi} B.,  {Barros} S.~C.~C.,  {Santos} N.~C.,  {Correia} A.~C.~M.,
  {Maxted} P.~F.~L.,  {Bou{\'e}} G.,   {Laskar} J.,  2019, \mn@doi [\aap]
  {10.1051/0004-6361/201834215}, \href
  {http://adsabs.harvard.edu/abs/2019A%26A...621A.117A} {621, A117}

\bibitem[\protect\citeauthoryear{Akinsanmi, Santos, Faria, Oshagh, Barros,
  Santerne  \& Charnoz}{Akinsanmi et~al.}{2020}]{akin20}
Akinsanmi B.,  Santos N.~C.,  Faria J.~P.,  Oshagh M.,  Barros S. C.~C.,
  Santerne A.,   Charnoz S.,  2020, \mn@doi [Astronomy {\&} Astrophysics]
  {10.1051/0004-6361/202037618}, 635, L8

\bibitem[\protect\citeauthoryear{Barnes \& Fortney}{Barnes \&
  Fortney}{2003}]{barnes03}
Barnes J.~W.,  Fortney J.~J.,  2003, The Astrophysical Journal, 588, 545

\bibitem[\protect\citeauthoryear{Barnes \& Fortney}{Barnes \&
  Fortney}{2004}]{barnes}
Barnes J.~W.,  Fortney J.~J.,  2004, The Astrophysical Journal, 616, 1193

\bibitem[\protect\citeauthoryear{Beichman et~al.,}{Beichman
  et~al.}{2014}]{beichman}
Beichman C.,  et~al., 2014, Publications of the Astronomical Society of the
  Pacific, 126, 1134

\bibitem[\protect\citeauthoryear{{Biersteker} \& {Schlichting}}{{Biersteker} \&
  {Schlichting}}{2017}]{biersteker}
{Biersteker} J.,  {Schlichting} H.,  2017, \mn@doi [\aj]
  {10.3847/1538-3881/aa88c2}, \href
  {http://adsabs.harvard.edu/abs/2017AJ....154..164B} {154, 164}

\bibitem[\protect\citeauthoryear{{Bouchy}, {Pepe}  \& {Queloz}}{{Bouchy}
  et~al.}{2001}]{bouchy}
{Bouchy} F.,  {Pepe} F.,   {Queloz} D.,  2001, \mn@doi [\aap]
  {10.1051/0004-6361:20010730}, \href
  {https://ui.adsabs.harvard.edu/abs/2001A&A...374..733B} {374, 733}

\bibitem[\protect\citeauthoryear{{Bouchy} et~al.,}{{Bouchy}
  et~al.}{2017}]{nirps}
{Bouchy} F.,  et~al., 2017, \mn@doi [The Messenger] {10.18727/0722-6691/5034},
  \href {https://ui.adsabs.harvard.edu/abs/2017Msngr.169...21B} {169, 21}

\bibitem[\protect\citeauthoryear{Carter \& Winn}{Carter \&
  Winn}{2010a}]{carterwinna}
Carter J.~A.,  Winn J.~N.,  2010a, The Astrophysical Journal, 709, 1219

\bibitem[\protect\citeauthoryear{Carter \& Winn}{Carter \&
  Winn}{2010b}]{carterwinnb}
Carter J.~A.,  Winn J.~N.,  2010b, The Astrophysical Journal, 716, 850

\bibitem[\protect\citeauthoryear{{Chaplin}, {Cegla}, {Watson}, {Davies}  \&
  {Ball}}{{Chaplin} et~al.}{2019}]{chaplin}
{Chaplin} W.~J.,  {Cegla} H.~M.,  {Watson} C.~A.,  {Davies} G.~R.,   {Ball}
  W.~H.,  2019, \mn@doi [\aj] {10.3847/1538-3881/ab0c01}, \href
  {https://ui.adsabs.harvard.edu/abs/2019AJ....157..163C} {157, 163}

\bibitem[\protect\citeauthoryear{{Chiavassa} et~al.,}{{Chiavassa}
  et~al.}{2017}]{chiavassa}
{Chiavassa} A.,  et~al., 2017, \mn@doi [A\&A] {10.1051/0004-6361/201528018},
  597, A94

\bibitem[\protect\citeauthoryear{{Claret} \& {Bloemen}}{{Claret} \&
  {Bloemen}}{2011}]{claret}
{Claret} A.,  {Bloemen} S.,  2011, \mn@doi [\aap]
  {10.1051/0004-6361/201116451}, \href
  {http://adsabs.harvard.edu/abs/2011A%26A...529A..75C} {529, A75}

\bibitem[\protect\citeauthoryear{{Correia}}{{Correia}}{2014}]{correia14}
{Correia} A. C.~M.,  2014, \mn@doi [\aap] {10.1051/0004-6361/201424733}, \href
  {https://ui.adsabs.harvard.edu/#abs/2014A&A...570L...5C} {570}

\bibitem[\protect\citeauthoryear{{Donati} et~al.,}{{Donati}
  et~al.}{2018}]{spirou}
{Donati} J.-F.,  et~al., 2018, {SPIRou: A NIR Spectropolarimeter/High-Precision
  Velocimeter for the CFHT}.
Springer International Publishing, p.~107,
  \mn@doi{10.1007/978-3-319-55333-7_107}

\bibitem[\protect\citeauthoryear{{Doyle}, {Davies}, {Smalley}, {Chaplin}  \&
  {Elsworth}}{{Doyle} et~al.}{2014}]{doyle14}
{Doyle} A.~P.,  {Davies} G.~R.,  {Smalley} B.,  {Chaplin} W.~J.,   {Elsworth}
  Y.,  2014, \mn@doi [\mnras] {10.1093/mnras/stu1692}, \href
  {https://ui.adsabs.harvard.edu/abs/2014MNRAS.444.3592D} {444, 3592}

\bibitem[\protect\citeauthoryear{{Faria} et~al.,}{{Faria}
  et~al.}{2020}]{faria19}
{Faria} J.~P.,  et~al., 2020, \mn@doi [\aap] {10.1051/0004-6361/201936389},
  \href {https://ui.adsabs.harvard.edu/abs/2020A&A...635A..13F} {635, A13}

\bibitem[\protect\citeauthoryear{{Foreman-Mackey}, {Hogg}, {Lang}  \&
  {Goodman}}{{Foreman-Mackey} et~al.}{2013}]{foreman}
{Foreman-Mackey} D.,  {Hogg} D.~W.,  {Lang} D.,   {Goodman} J.,  2013, \mn@doi
  [\pasp] {10.1086/670067}, \href
  {http://adsabs.harvard.edu/abs/2013PASP..125..306F} {125, 306}

\bibitem[\protect\citeauthoryear{Foreman-Mackey, Agol, Ambikasaran  \&
  Angus}{Foreman-Mackey et~al.}{2017}]{dfm17}
Foreman-Mackey D.,  Agol E.,  Ambikasaran S.,   Angus R.,  2017, The
  Astronomical Journal, 154, 220

\bibitem[\protect\citeauthoryear{Gaudi \& Winn}{Gaudi \&
  Winn}{2007}]{gaudi_winn}
Gaudi B.~S.,  Winn J.~N.,  2007, \mn@doi [The Astrophysical Journal]
  {10.1086/509910}, 655, 550

\bibitem[\protect\citeauthoryear{{Guillot}, {Burrows}, {Hubbard}, {Lunine}  \&
  {Saumon}}{{Guillot} et~al.}{1996}]{guillot}
{Guillot} T.,  {Burrows} A.,  {Hubbard} W.~B.,  {Lunine} J.~I.,   {Saumon} D.,
  1996, \mn@doi [\apjl] {10.1086/309935}, \href
  {http://adsabs.harvard.edu/abs/1996ApJ...459L..35G} {459, L35}

\bibitem[\protect\citeauthoryear{{Hellard}, {Csizmadia}, {Padovan}, {Rauer},
  {Cabrera}, {Sohl}, {Spohn}  \& {Breuer}}{{Hellard} et~al.}{2019}]{hellard19}
{Hellard} H.,  {Csizmadia} S.,  {Padovan} S.,  {Rauer} H.,  {Cabrera} J.,
  {Sohl} F.,  {Spohn} T.,   {Breuer} D.,  2019, \mn@doi [\apj]
  {10.3847/1538-4357/ab2048}, \href
  {https://ui.adsabs.harvard.edu/abs/2019ApJ...878..119H} {878, 119}

\bibitem[\protect\citeauthoryear{Kaspi \& Showman}{Kaspi \&
  Showman}{2015}]{Kaspi_2015}
Kaspi Y.,  Showman A.~P.,  2015, \mn@doi [The Astrophysical Journal]
  {10.1088/0004-637x/804/1/60}, 804, 60

\bibitem[\protect\citeauthoryear{{Kipping}}{{Kipping}}{2013}]{kipping13}
{Kipping} D.~M.,  2013, \mn@doi [\mnras] {10.1093/mnras/stt1435}, \href
  {http://adsabs.harvard.edu/abs/2013MNRAS.435.2152K} {435, 2152}

\bibitem[\protect\citeauthoryear{{Laskar} \& {Robutel}}{{Laskar} \&
  {Robutel}}{1993}]{laskar_robutel}
{Laskar} J.,  {Robutel} P.,  1993, \mn@doi [\nat] {10.1038/361608a0}, \href
  {https://ui.adsabs.harvard.edu/abs/1993Natur.361..608L} {361, 608}

\bibitem[\protect\citeauthoryear{{Li} \& {Lai}}{{Li} \& {Lai}}{2020}]{li_lai}
{Li} J.,  {Lai} D.,  2020, arXiv e-prints, \href
  {https://ui.adsabs.harvard.edu/abs/2020arXiv200507718L} {p. arXiv:2005.07718}

\bibitem[\protect\citeauthoryear{{Lissauer}}{{Lissauer}}{1995}]{lissauer}
{Lissauer} J.~J.,  1995, \mn@doi [\icarus] {10.1006/icar.1995.1057}, \href
  {https://ui.adsabs.harvard.edu/abs/1995Icar..114..217L} {114, 217}

\bibitem[\protect\citeauthoryear{{McLaughlin}}{{McLaughlin}}{1924}]{mclaughlin}
{McLaughlin} D.~B.,  1924, \mn@doi [\apj] {10.1086/142826}, \href
  {http://adsabs.harvard.edu/abs/1924ApJ....60...22M} {60}

\bibitem[\protect\citeauthoryear{{Ohta}, {Taruya}  \& {Suto}}{{Ohta}
  et~al.}{2009}]{ohta}
{Ohta} Y.,  {Taruya} A.,   {Suto} Y.,  2009, \mn@doi [\apj]
  {10.1088/0004-637X/690/1/1}, \href
  {http://adsabs.harvard.edu/abs/2009ApJ...690....1O} {690, 1}

\bibitem[\protect\citeauthoryear{Oshagh}{Oshagh}{2018a}]{oshagh17}
Oshagh M.,  2018a, in Campante T.~L.,  Santos N.~C.,   Monteiro M. J. P. F.~G.,
   eds, Asteroseismology and Exoplanets: Listening to the Stars and Searching
  for New Worlds. Springer Publishing, Cham, pp 239--249

\bibitem[\protect\citeauthoryear{{Oshagh}}{{Oshagh}}{2018b}]{oshagh18}
{Oshagh} M.,  2018b, \mn@doi [Asteroseismology and Exoplanets: Listening to the
  Stars and Searching for New Worlds] {10.1007/978-3-319-59315-9_13}, \href
  {http://adsabs.harvard.edu/abs/2018ASSP...49..239O} {49, 239}

\bibitem[\protect\citeauthoryear{{Oshagh}, {Santos}, {Boisse}, {Bou{\'e}},
  {Montalto}, {Dumusque}  \& {Haghighipour}}{{Oshagh}
  et~al.}{2013}]{oshagh2013}
{Oshagh} M.,  {Santos} N.~C.,  {Boisse} I.,  {Bou{\'e}} G.,  {Montalto} M.,
  {Dumusque} X.,   {Haghighipour} N.,  2013, \mn@doi [\aap]
  {10.1051/0004-6361/201321309}, \href
  {http://adsabs.harvard.edu/abs/2013A%26A...556A..19O} {556, A19}

\bibitem[\protect\citeauthoryear{{Peale}}{{Peale}}{1999}]{peale}
{Peale} S.~J.,  1999, \mn@doi [\araa] {10.1146/annurev.astro.37.1.533}, \href
  {http://adsabs.harvard.edu/abs/1999ARA%26A..37..533P} {37, 533}

\bibitem[\protect\citeauthoryear{Pepe et~al.,}{Pepe et~al.}{2014}]{pepe}
Pepe F.,  et~al., 2014, \mn@doi [Astronomische Nachrichten]
  {10.1002/asna.201312004}, 335, 8

\bibitem[\protect\citeauthoryear{{Quirrenbach} et~al.,}{{Quirrenbach}
  et~al.}{2016}]{carmenes}
{Quirrenbach} A.,  et~al., 2016, {CARMENES: an overview six months after first
  light}.
SPIE, p. 990812, \mn@doi{10.1117/12.2231880}

\bibitem[\protect\citeauthoryear{{Rauer} et~al.,}{{Rauer} et~al.}{2014}]{rauer}
{Rauer} H.,  et~al., 2014, \mn@doi [Experimental Astronomy]
  {10.1007/s10686-014-9383-4}, \href
  {http://adsabs.harvard.edu/abs/2014ExA....38..249R} {38, 249}

\bibitem[\protect\citeauthoryear{{Rossiter}}{{Rossiter}}{1924}]{rossiter}
{Rossiter} R.~A.,  1924, \mn@doi [\apj] {10.1086/142825}, \href
  {http://adsabs.harvard.edu/abs/1924ApJ....60...15R} {60}

\bibitem[\protect\citeauthoryear{Seager \& Hui}{Seager \&
  Hui}{2002}]{seagerhui}
Seager S.,  Hui L.,  2002, The Astrophysical Journal, 574, 1004

\bibitem[\protect\citeauthoryear{{Torres}, {Winn}  \& {Holman}}{{Torres}
  et~al.}{2008}]{torres}
{Torres} G.,  {Winn} J.~N.,   {Holman} M.~J.,  2008, \mn@doi [\apj]
  {10.1086/529429}, \href {http://adsabs.harvard.edu/abs/2008ApJ...677.1324T}
  {677, 1324}

\bibitem[\protect\citeauthoryear{{Triaud} et~al.,}{{Triaud}
  et~al.}{2009}]{triaud09}
{Triaud} A.~H.~M.~J.,  et~al., 2009, \mn@doi [\aap]
  {10.1051/0004-6361/200911897}, \href
  {https://ui.adsabs.harvard.edu/abs/2009A&A...506..377T} {506, 377}

\bibitem[\protect\citeauthoryear{{Wyttenbach}, {Ehrenreich}, {Lovis}, {Udry}
  \& {Pepe}}{{Wyttenbach} et~al.}{2015}]{wyttenbach}
{Wyttenbach} A.,  {Ehrenreich} D.,  {Lovis} C.,  {Udry} S.,   {Pepe} F.,  2015,
  \mn@doi [\aap] {10.1051/0004-6361/201525729}, \href
  {https://ui.adsabs.harvard.edu/abs/2015A&A...577A..62W} {577, A62}

\bibitem[\protect\citeauthoryear{Zhu, Huang, Zhou  \& Lin}{Zhu
  et~al.}{2014}]{zhu}
Zhu W.,  Huang C.~X.,  Zhou G.,   Lin D. N.~C.,  2014, ApJ, 796, 67

\bibitem[\protect\citeauthoryear{Zuluaga, Kipping, Sucerquia  \&
  Alvarado}{Zuluaga et~al.}{2015}]{zuluaga}
Zuluaga J.~I.,  Kipping D.~M.,  Sucerquia M.,   Alvarado J.~A.,  2015, The
  Astrophysical Journal Letters, 803, L14

\makeatother
\end{thebibliography}



\newpage
\clearpage
\appendix

\section{Additional Figures}

\begin{figure}
	\includegraphics[width=0.9\columnwidth]{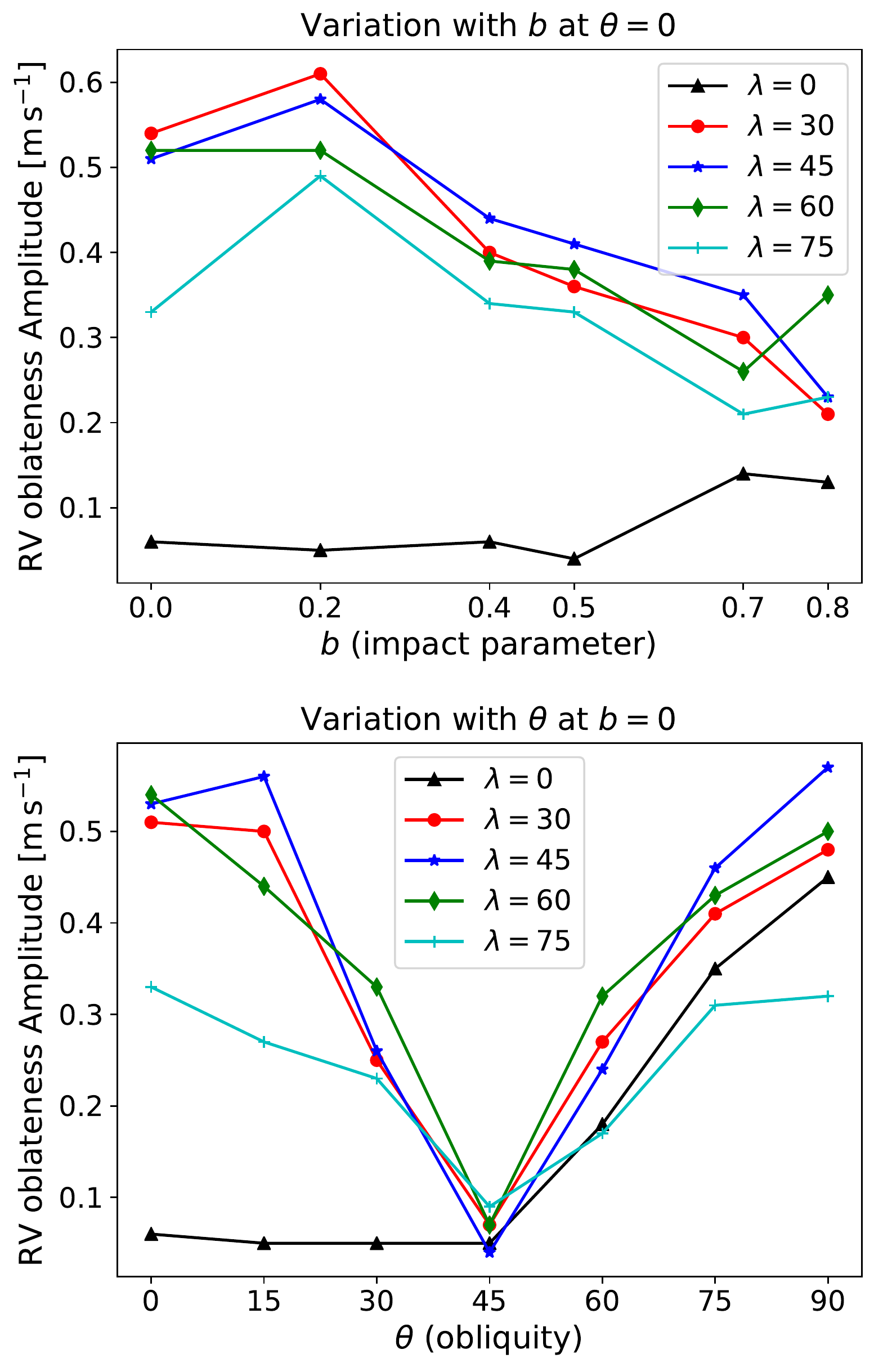}
	\caption{Variation of RV oblateness amplitude at different values of spin-orbit angle ($\lambda$). Top: Amplitude variation with $b$ at $\theta=0$. Bottom: Amplitude variation with $\theta$ at $b=0$. }
    \label{lbdafits}
\end{figure}

\newpage



\bsp	
\label{lastpage}
\end{document}